\documentclass[aps,pra,preprint,superscriptaddress, showpacs]{revtex4-2}
\usepackage{graphicx}
\usepackage{natbib}
\newcommand*{\Scale}[2][4]{\scalebox{#1}{$#2$}}%
\newcommand*{\Resize}[2]{\resizebox{#1}{!}{$#2$}}%

\begin{document}

\title{Influence of squeezing on the weak-to-strong measurement transition}

\author{Kevin Araya-Sossa}
\email{kjaraya@uc.cl}
\affiliation{
	Instituto de Física, Pontificia Universidad Católica de Chile, Casilla 306, Santiago,
	Chile}
\author{Miguel Orszag}
\email{Corresponding author: miguel.orszag@umayor.cl}
\affiliation{
	Instituto de Física, Pontificia Universidad Católica de Chile, Casilla 306, Santiago,
	Chile}
\affiliation{Centro de Óptica e Información Cuántica, Universidad Mayor, Camino la Pirámide 5750, Huechuraba, Santiago, Chile}

\date{\today}

\begin{abstract}

In this work, we study the measurement transition for a coherent squeezed pointer state through a transition factor $\Gamma$ that involves a system-pointer coupling by using an arbitrary measured observable $A$. In addition, we show that the shift in the pointer's position and momentum establishes a relationship with a new value defined as the transition value, which generalizes the weak value as well as the conditional expectaction value. Furthermore, a new strategy is introduced to achieve different measurement regimes by just adjusting the $r$ and $\phi_{\xi}$ parameters of the coherent squeezed pointer state, opening an interesting way to test quantum mechanics foundations. Our scheme has been theoretically applied in a trapped ion iluminated by a bichromatic laser beam, with a high potential to be implemented in future experimental setups.

\end{abstract}


\maketitle

\section{\label{sec:intro}Introduction}

Measurements in quantum mechanics have always been an essential challenge to understand a variety of physical phenomena \cite{peres1993quantum,aharonov2005quantum,jacobs2014quantum}. A simple model which describes strong quantum measurements was developed by von Neumann \cite{neumann1955mathematical} and completely generalized by Ozawa some years later \cite{ozawa1984quantum}. This model involves the coupling between two quantum systems namely the measured system and the pointer or meter, controlled by a coupling constant.   

From this model, Aharanov and co-workers \cite{aharonov1988result} proposed to extend the strong measurement to a weak regime with the help of the time-symmetry theory of quantum mechanics \cite{aharonov1964time}. By considering a gaussian pointer state as well as pre-selection and post-selection of the system state, through the readout of the pointer, we get the weak value
\begin{equation}\label{weak.value}
A_{w}=\frac{\langle F|A|I\rangle}{\langle F|I\rangle},
\end{equation}
where $|I\rangle$ and $|F\rangle$ are the pre-selected and post-selected state, respectively. $A$ is the measured observable with eigenvalues $a_j$ and eigenstates $|a_{j}\rangle$. A useful physical phenomena can be observed and amplified if a suitable post-selected state is chosen to be almost orthogonal to the pre-selected state. Such an interesting quantum effect is the so-called weak value amplification (WVA) and it has several applications as a technique to amplify very small signals for observing and studying quantum physical effects like the spin Hall effect \cite{hosten2008observation}, the deflection of a light beam \cite{dixon2009ultrasensitive}, velocity displacement \cite{viza2013weak} and temperature shift \cite{egan2012weak}, among others \cite{starling2010continuous, simon2011fock,magana2014amplification,jordan2014technical,zhang2015application}. Nevertheless, having an anormalous weak value means a small post-selection probability $P_{post}=|\langle F|I \rangle|^2$.

An important feature of the weak value is that it can be a complex number in contrast to the expectaction value
\begin{equation}\label{strong.value}
A_{s}=\frac{\langle F|A|F\rangle}{\langle F|F\rangle},
\end{equation}
which coincide with the weak value when the state $|I\rangle$ is parallel to the state $|F\rangle$ or if $|F\rangle$ is eigenstate of $A$. The physical interpretation of this property was derived by Josza \cite{jozsa2007complex}, where the shifts of the pointer have a direct relation with the imaginary and real part of this value in the weak regime.


Since the measurement problem is still under study, a fundamental research on the foundations of quantum mechanics is related to the weak to strong measurement transition \cite{ferraioli2019measurement,gyongyosi2019dense}. Firstly, Zhu \textit{et al.} \cite{zhu2011quantum} studied the quantum measurement transition by considering extreme conditions of the system-pointer coupling. Specifically, the transition from the weakest to the strongest regime was found, for all cases, and in particular for the two extremes by using a gaussian wavepacket as pointer. These results allowed to extend the Josza's theorem for the strong regime by relating the displacements in the measurement pointer with the conditional expectaction value [Aharonov-Bergmann-Lebowitz (ABL) rule] \cite{aharonov1964time, aharonov2008two}
\begin{equation}\label{conditional.value}
A_{c}=\sum_{j} a_{j}\,\frac{\left|\langle F|a_{j} \rangle\langle a_{j}|I \rangle\right|^2}{\sum_{k}\left|\langle F|a_{k} \rangle\langle a_{k}|I \rangle\right|^2}.
\end{equation}
Here, the conditional concept arises from the post-selection process. Next, Ban in Ref. \cite{ban2015conditional} provided the conditional average as a combination of the weak value and the conditional expectaction value to find a general average value. Finally, the measurement transition was experimentally investigated by modulating a global transition factor $\Gamma=gt/X_{0}$ where the time of interaction $t$ between the system and the pointer, the coupling constant $g$ and the width of a initial gaussian pointer $X_{0}$ can be modified \cite{pan2020weak}. To test it, a simple experimental setup involving a single trapped ${}^{40}\text{Ca}^{+}$ ion irradiated with a bichromatic beam light was used \cite{wu2019scheme}. 

On the other hand, Turek and his collaborators explored advantages of implementing nonclassical pointer states theoretically in the measurement transition for a measured observable satisfying the property $A^2=I$ \cite{turek2015advantages}. Particularly, squeezed states have been sucessfully generated via motional states of a ${}^{40}\text{Ca}^{+}$ ion in a trap through different methods. Among them, Kienzler \textit{et al.} demonstrated the generation of squeezing produced experimentally by reservoir engineering \cite{kienzler2015quantum}. More recently, Drechsler and co-workers introduced a new method to squeeze the motion of the trapped ion by placing the ion inside a time-varying potential controlled by the phase of an optical lattice \cite{drechsler2020state}. 
 


Inspired by current theoretical and experimental work, we generalize the Josza's theorem and find the general position and momentum average of a coherent squeezed pointer state \cite{caves1980quantum,caves1981quantum}, when measuring an arbitrary observable $A$, expressed in terms of a newly defined transition value $A_{T}$. We demonstrated that:
$
\left(A_{T}\right)_{\Gamma \rightarrow 0}=A_{w}\, \; \text{and}\; \left(A_{T}\right)_{\Gamma \rightarrow \infty}=A_{c}, 
$
unifying the definitions of the weak value and the conditional expectacion value without considering a mixture of them. We also show that one can go from the weak to the strong regime by varying the squeezing parameters and fixing the global transition factor, as an alternative way to study the measurement transition. To carry out our proposal, we apply all these ideas to the ${}^{40}\text{Ca}^{+}$ ion stored inside the Paul trap interacting with two laser fields at specific frequencies. 



Our work is organized as followed. In Sec. \ref{sec:formalism} we extend the idea originally proposed by Josza for coherent squeezed states and in Sec. \ref{sec:formalism2} we determine the shifts of the pointer for extreme regimes by controlling the global transition factor. Next, in Sec. \ref{sec:formalism3} we analize the effect of the squeezing paramenters on the weak to strong measurement transition. After that, in Sec. \ref{sec:formalism4} our strategy is applied to the case of a trapped ion in a bichromatic field. Finally, we discuss the impact of our results and a possible experimental application.


\section{\label{sec:formalism}Generalized Josza's theorem}

Let us start for considering the standard formalism of quantum measurement \cite{neumann1955mathematical}, where the system interacts with the pointer through the following coupling hamiltonian
\begin{equation}\label{hamiltonian}
H= g\, A \otimes P,
\end{equation}
$A$ and $P$ being the measured observable and the momentum of the pointer, respectively. Here, $P$ can be written in terms of annihilation and creation operators as follows
\begin{equation}\label{momentum.operator}
P=\frac{\hbar}{2 X_{0}}i\left(a^{\dagger}-a\right),
\end{equation}
where $X_{0}=\sqrt{\hbar/2m\nu}$ is the size of the gaussian ground state that depends on the mass of the pointer $m$ and the frequency $\nu$ with which the system oscillates. Note that the hamiltonian contains a coupling constant $g$ responsible of the interaction between both systems.

We now assume an initial system-pointer state of the form:
\begin{equation}
|\Psi_{in}\rangle=|I\rangle\otimes |\phi_{in}\rangle, 
\end{equation}
where
\begin{equation}
|I\rangle=\sum_{j} \alpha_{j}|a_{j}\rangle \quad \text{and} \quad |\phi_{in}\rangle=|\alpha, \xi\rangle
\end{equation}
are the initial states of the system and the pointer, respectively. We take as initial pointer, the coherent squeezed state \cite{caves1980quantum,caves1981quantum} 
\begin{equation}\label{squezeed.state}
|\alpha, \xi\rangle=D(\alpha)S(\xi)|0\rangle.
\end{equation}
Here, 
\begin{equation}
D(\alpha)=\textrm{exp}\left(\alpha a^{\dagger}-\alpha^{*}a\right) \quad \text{and} \quad S(\xi)=\textrm{exp}\left(\frac{1}{2}\xi^{*}a^{2}-\frac{1}{2}\xi a^{\dagger 2}\right)
\end{equation}
are the displacement operator and the squeezing operator \cite{glauber1963coherent,gerry2005introductory}, respectively, with $\alpha=|\alpha|\, \textrm{exp}(i\phi_{\alpha})$ and $\xi=r\,\textrm{exp}(i\phi_{\xi})$.
Then the joint system evolves by means of the hamiltonian [See Eq. (\ref{hamiltonian})] as
\begin{eqnarray}\label{evolved.state}
|\Psi_{evol}\rangle &=&\textrm{exp}\left(-\frac{i}{\hbar}\int_{0}^{t} H(\tau)\,d\tau\right)\,|\Psi_{in}\rangle \nonumber \\
&=&\sum_{j}\alpha_{j}\,\textrm{exp}\left(-\frac{i}{\hbar}ga_{j}t P\right)\,|a_{j}\rangle\otimes D(\alpha)S(\xi)|0\rangle\nonumber\\
&=&\sum_{j} \alpha_{j}\,|a_{j}\rangle\otimes D\left(\frac{\Gamma}{2}a_{j} \right)D(\alpha)S(\xi)|0\rangle,
\end{eqnarray}
where $\Gamma=gt/X_{0}$ is the transition measurement factor. By using the following property \cite{gerry2005introductory}
\begin{equation}\label{coherent.property}
D(y)D(z)=\textrm{exp}\left(\frac{1}{2}yz^{*}-\frac{1}{2}y^{*}z\right) D(y+z),
\end{equation}
Eq. (\ref{evolved.state}) can be re-written as
\begin{eqnarray}
|\Psi_{evol}\rangle&=&\sum_{j}\alpha_{j}\textrm{exp}\left(-i \frac{\Gamma}{2}a_{j}\,\textrm{Im}(\alpha)\right)|a_{j}\rangle\otimes  D\left(\frac{\Gamma}{2}a_{j}+\alpha\right)S(\xi)|0\rangle \nonumber\\
&=&\sum_{j}\alpha_{j}\textrm{exp}\left(-i \frac{\Gamma}{2}a_{j}\,\textrm{Im}(\alpha)\right)|a_{j}\rangle\otimes\left|\frac{\Gamma}{2}a_{j}+\alpha, \xi\right\rangle.
\end{eqnarray}
Finally, by post-selecting the system state $|F\rangle=\sum_{k}\beta_{k}\, |a_{k}\rangle$, we get a final pointer state
\begin{eqnarray}\label{final.pointer}
|\phi_{fin}\rangle&=&\langle F|\sum_{j}\alpha_{j}\textrm{exp}\left(-i \frac{\Gamma}{2}a_{j}\,\textrm{Im}(\alpha)\right)|a_{j}\rangle\otimes\left|\frac{\Gamma}{2}a_{j}+\alpha, \xi\right\rangle \nonumber\\
&=&\sum_{j} \alpha_{j}\beta_{j}^{*}\,\textrm{exp}\left(-i \frac{\Gamma}{2}a_{j}\,\textrm{Im}(\alpha)\right)\left|\frac{\Gamma}{2}a_{j}+\alpha, \xi\right\rangle.
\end{eqnarray}

On the other hand, we define the transition value as 
\begin{equation}\label{transition.value}
A_{T}=\frac{\langle \phi_{fin}|\phi_{fin}^{A}\rangle}{\langle \phi_{fin}|\phi_{fin} \rangle},
\end{equation}
with
\begin{equation}
|\phi_{fin}^{A}\rangle=\langle F|A|\Psi_{evol}\rangle.
\end{equation}
This general value is introduced to extend the values that represent the extreme regimes of the quantum measurement transition. In particular, for the coherent squeezed pointer state, the transition value takes the form
\begin{equation}\label{transition.value.squezeed.state}
A_{T}=\frac{\sum_{j,k}\alpha_{j}\beta_{j}^{*}\alpha_{k}^{*}\beta_{k}a_{j}\,\textrm{exp}\left[-i\Gamma\,\textrm{Im}(\alpha)\left(a_{j}-a_{k}\right)\right]\,\textrm{exp}\left[-\frac{\Gamma^{2}}{8}\left(a_{j}-a_{k}\right)^{2}|\mu+\nu|^2\right]}{\sum_{j,k}\alpha_{j}\beta_{j}^{*}\alpha_{k}^{*}\beta_{k}\,\textrm{exp}\left[-i\Gamma\,\textrm{Im}(\alpha)\left(a_{j}-a_{k}\right)\right]\,\textrm{exp}\left[-\frac{\Gamma^{2}}{8}\left(a_{j}-a_{k}\right)^{2}|\mu+\nu|^2\right]},
\end{equation}
where $\mu=\cosh{r}$ and $\nu=\sinh{r}\,\textrm{exp}(i\phi_{\xi})$. Here, we used the property in Eq. (\ref{inner.product.squeezed.state}).

From Eq. (\ref{final.pointer}), it is straightforward to show that the shift in the pointer's position after post-selection is
\begin{eqnarray}\label{position.shift}
\delta x &=& \frac{\langle \phi_{fin}|X|\phi_{fin} \rangle}{\langle \phi_{fin}|\phi_{fin}\rangle}-\langle \phi_{in}|X|\phi_{in} \rangle \nonumber \\
&=&\Resize{13cm}{\frac{gt}{2}\frac{\sum_{j,k}\alpha_{j}\beta_{j}^{*}\alpha_{k}^{*}\beta_{k}\left(a_{k}+a_{j}\right)\,\Scale[0.82]{\textrm{exp}}\left[-i\Gamma\,\Scale[0.82]{\textrm{Im}}(\alpha)\left(a_{j}-a_{k}\right)\right]\,\Scale[0.82]{\textrm{exp}}\left[-\frac{\Gamma^{2}}{8}\left(a_{j}-a_{k}\right)^{2}|\mu+\nu|^2\right]}{\sum_{j,k}\alpha_{j}\beta_{j}^{*}\alpha_{k}^{*}\beta_{k}\,\Scale[0.85]{\textrm{exp}}\left[-i\Gamma\,\Scale[0.85]{\textrm{Im}}(\alpha)\left(a_{j}-a_{k}\right)\right]\,\Scale[0.85]{\textrm{exp}}\left[-\frac{\Gamma^{2}}{8}\left(a_{j}-a_{k}\right)^{2}|\mu+\nu|^2\right]}} \nonumber \\
&&  \Resize{13cm}{-igt\mu\Scale[0.85]{\textrm{Im}}(\nu)\frac{\sum_{j,k}\alpha_{j}\beta_{j}^{*}\alpha_{k}^{*}\beta_{k}\left(a_{k}-a_{j}\right)\,\Scale[0.82]{\textrm{exp}}\left[-i\Gamma\,\Scale[0.82]{\textrm{Im}}(\alpha)\left(a_{j}-a_{k}\right)\right]\,\Scale[0.82]{\textrm{exp}}\left[-\frac{\Gamma^{2}}{8}\left(a_{j}-a_{k}\right)^{2}|\mu+\nu|^2\right]}{\sum_{j,k}\alpha_{j}\beta_{j}^{*}\alpha_{k}^{*}\beta_{k}\,\Scale[0.82]{\textrm{exp}}\left[-i\Gamma\,\Scale[0.82]{\textrm{Im}}(\alpha)\left(a_{j}-a_{k}\right)\right]\,\Scale[0.82]{\textrm{exp}}\left[-\frac{\Gamma^{2}}{8}\left(a_{j}-a_{k}\right)^{2}|\mu+\nu|^2\right]}}\nonumber \\
&=& gt \,\textrm{Re}(A_{T})-2gt\mu\,\textrm{Im}(\nu)\,\textrm{Im}(A_{T}).
\end{eqnarray}
Following the similar procedure, the momentum displacement in the pointer results in
\begin{eqnarray}\label{momentum.shift}
\delta p &=& \frac{\langle \phi_{fin}|P|\phi_{fin} \rangle}{\langle \phi_{fin}|\phi_{fin}\rangle}-\langle \phi_{in}|P|\phi_{in}\rangle \nonumber \\
&=&\Resize{13cm}{i\frac{gt}{\hbar}\,\Scale[0.85]{\textrm{Var}(P)_{in}}\frac{\sum_{j,k}\alpha_{j}\beta_{j}^{*}\alpha_{k}^{*}\beta_{k}\left(a_{k}-a_{j}\right)\,\Scale[0.82]{\textrm{exp}}\left[-i\Gamma\,\Scale[0.82]{\textrm{Im}}(\alpha)\left(a_{j}-a_{k}\right)\right]\,\Scale[0.82]{\textrm{exp}}\left[-\frac{\Gamma^{2}}{8}\left(a_{j}-a_{k}\right)^{2}|\mu+\nu|^2\right]}{\sum_{j,k}\alpha_{j}\beta_{j}^{*}\alpha_{k}^{*}\beta_{k}\,\Scale[0.82]{\textrm{exp}}\left[-i\Gamma\,\Scale[0.82]{\textrm{Im}}(\alpha)\left(a_{j}-a_{k}\right)\right]\,\Scale[0.82]{\textrm{exp}}\left[-\frac{\Gamma^{2}}{8}\left(a_{j}-a_{k}\right)^{2}|\mu+\nu|^2\right]}} \nonumber \\
&=& \frac{2gt}{\hbar}\,\textrm{Var}(P)_{in}\, \textrm{Im}(A_{T}),
\end{eqnarray}
where $\textrm{Var}(P)_{in}=\hbar^2|\mu+\nu|^2/4X_{0}^{2}$ is the initial variance in the pointer's momentum. In order to find the expressions shown in Eqs. (\ref{position.shift}) and (\ref{momentum.shift}), we used the following property [See Eq. (\ref{special.inner.product})] 
\begin{equation}
\left\langle \frac{\Gamma}{2}a_{k}+\alpha, \xi\right|a\left|\frac{\Gamma}{2}a_{j}+\alpha, \xi\right\rangle=\frac{\Gamma}{2}a_{j}\mu(\mu+\nu)-\frac{\Gamma}{2}a_{k}\nu(\mu+\nu^{*})+\alpha.
\end{equation}
It is important to highlight the fact that both displacements are related to the transition factor $\Gamma$ that determine the limits in the quantum measurement. In the following section, we analize their behavior in the weak and strong regime in details.

\section{\label{sec:formalism2}Limiting values of the pointer's shift}


In an effort to generalize the Josza's theorem \cite{jozsa2007complex}, we obtain the weak value and the conditional expectation value of the observable $A$. By applying Eqs. (\ref{weak.value}) and (\ref{conditional.value}), it is simple to show that
\begin{equation}\label{weak.and.strong.value}
A_{w}=\frac{\sum_{j}\alpha_{j}\beta_{j}^{*}a_{j}}{\sum_{j}\alpha_{j}\beta_{j}^{*}a_{j}} \quad \text{and} \quad A_{c}=\frac{\sum_{j} |a_{j}\beta_{j}^{*}|^{2} a_{j}}{\sum_{j} |a_{j}\beta_{j}^{*}|^{2}}.
\end{equation}
Without considering the parameters associated to squeezing, the measurement regime is determined by the strength of the dimensionless factor $\Gamma$ \cite{pan2020weak}. Thus, we can control the measurement regime by means of the parameters $g$, $t$ and $X_{0}$. Taking the weak limit, $\Gamma\rightarrow 0$ and Eq. (\ref{transition.value.squezeed.state}), the transition value takes the form 
\begin{eqnarray}
(A_{T})_{\Gamma \rightarrow 0}&=&\frac{\sum_{j, k}\alpha_{j}\beta_{j}^{*}\alpha_{k}^{*}\beta_{k}a_{j}}{\sum_{j, k}\alpha_{j}\beta_{j}^{*}\alpha_{k}^{*}\beta_{k}}\nonumber \\
&=&\frac{\sum_{j}\alpha_{j}\beta_{j}^{*}a_{j}}{\sum_{j}\alpha_{j}\beta_{j}^{*}}\nonumber \\
&=&A_{w}.
\end{eqnarray}
In contrast to the weak regime, in the strong regime, the parameter $\Gamma \rightarrow \infty$. Under this condition, the transition value becomes
\begin{eqnarray}
(A_{T})_{\Gamma \rightarrow \infty}&=&\Resize{13cm}{\lim_{\Gamma\rightarrow \infty}\frac{\sum_{j}|\alpha_{j}\beta_{j}^{*}|^2 a_{j}+\sum_{j \neq k}\alpha_{j}\beta_{j}^{*}\alpha_{k}^{*}\beta_{k}a_{j}\,\Scale[0.82]{\textrm{exp}}\left[-i\Gamma\,\Scale[0.82]{\textrm{Im}}(\alpha)\left(a_{j}-a_{k}\right)\right]\,\Scale[0.82]{\textrm{exp}}\left[ -\frac{\Gamma^2}{8}\left(a_{j}-a_{k}\right)^2|\mu+\nu|^2\right]}{\sum_{j}|\alpha_{j}\beta_{j}^{*}|^2+\sum_{j\neq k}\alpha_{j}\beta_{j}^{*}\alpha_{k}^{*}\beta_{k}\,\Scale[0.82]{\textrm{exp}}\left[-i\Gamma\,\Scale[0.82]{\textrm{Im}}(\alpha)\left(a_{j}-a_{k}\right)\right]\,\Scale[0.82]{\textrm{exp}}\left[ -\frac{\Gamma^2}{8}\left(a_{j}-a_{k}\right)^2|\mu+\nu|^2\right]}}\nonumber \\
&=&\frac{\sum_{j}|\alpha_{j}\beta_{j}^{*}|^2 a_{j}}{\sum_{j}|\alpha_{j}\beta_{j}^{*}|^2} \nonumber \\
&=&A_{c}. 
\end{eqnarray}
Clearly, we observe that the weak value and the conditional expectaction value establish a relationship with the weak and strong regime, respectively. This correspondence reveals the nature of the quantum measurement. For the weak regime, the shifts in the pointer's position and momentum [See Eqs (\ref{position.shift}) and (\ref{momentum.shift})] behave as
\begin{equation}\label{weak.shift}
(\delta x)_{\Gamma\rightarrow 0}=gt \,\textrm{Re}(A_{w})-2gt\mu\,\textrm{Im}(\nu)\,\textrm{Im}(A_{w})\quad \text{and} \quad  (\delta p)_{\Gamma\rightarrow 0}=\frac{2gt}{\hbar}\,\textrm{Var}(P)_{in}\, \textrm{Im}(A_{w}).
\end{equation}
While in the strong regime
\begin{equation}\label{strong.shift}
(\delta x)_{\Gamma\rightarrow \infty}=gt \,\textrm{Re}(A_{c})\quad \text{and} \quad (\delta p)_{\Gamma\rightarrow \infty}=0.
\end{equation}
Since the conditional expectation value is real, the displacement in the pointer's momentum vanishes. 

For the coherent pointer state $(r=0)$, both displacements in the weak regime are connected with the real and imaginary part of the weak value, recovering the results obtained by Josza \cite{jozsa2007complex}. Similarly, in the strong regime, the real and imaginary part of the conditional expectaction value is in accordance with the shift of the pointer's position and momentum, which give us an extension of the Josza's theorem. In particular, Zhu \textit{et. al} \cite{zhu2011quantum} reported this important result for the pointer's ground state $(\alpha=0)$.

Until now, we have only considered the transition factor $\Gamma$. In order to analize the effects of squeezing on the pointer's displacements, we will study how squeezing parameters influence the measurement transition.

\section{\label{sec:formalism3}The effect of the squeezing on the measurement transition}

\begin{figure}
	\centering
	\includegraphics[width=8.6cm, height=12.0cm]{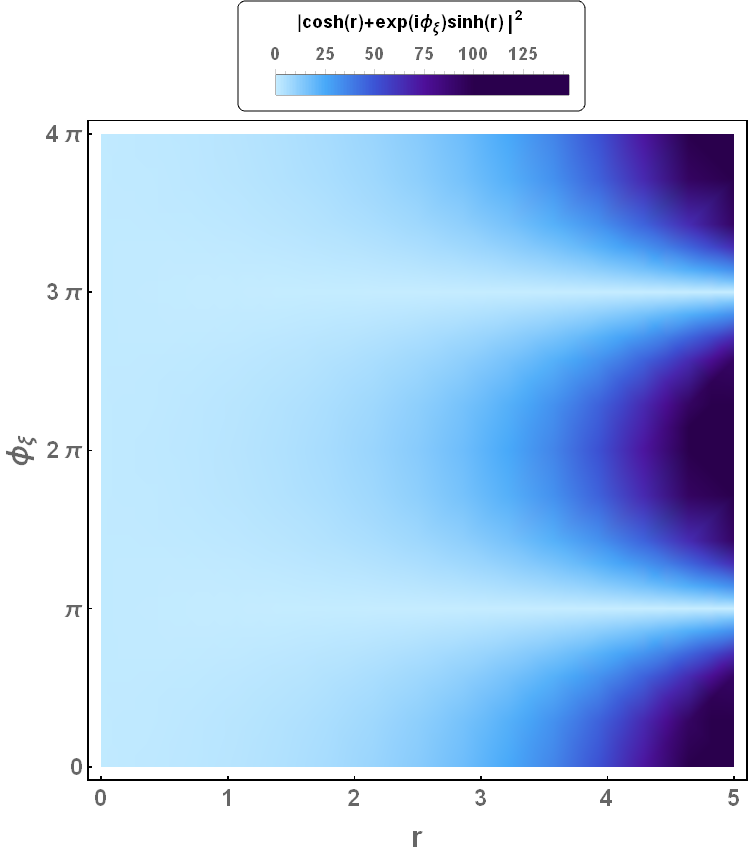}
	\caption{(Color online) The amount $|\mu+\nu|^2$ as a function of the squeezing parameters $r$ and $\phi_{\xi}$.}
	\label{fig1}
\end{figure}

Evidently, the transition value behaves as the weak value or the conditional expectaction value modulated by the terms on the exponential function in Eq. (\ref{transition.value.squezeed.state}). By setting the transition factor $\Gamma$ and varying the squeezing parameters $\left(\alpha, r \; \text{and} \; \phi_{\xi}\right)$, it is possible to reach the measurement transition. Specifically, this transition depends on the terms $\textrm{Im}(\alpha)$ and $|\mu+\nu|^2$. Hence, by minimizing and maximizing these quantities, we achieve our goal. Clearly, the first term is easier to optimize than the other one because it is linear in $\alpha$. The second term can be re-written as
\begin{equation}
|\mu+\nu|^2= \left|\cosh{r}+\sinh{r}\,\textrm{exp}\left(i\phi_{\xi}\right)\right|^2=\cosh{(2r)}+\sinh{(2r)}\cos{(\phi_{\xi})},
\end{equation}
which is dependent on the parameters $r$ and $\phi_{\xi}$. 
More specifically
\begin{equation}
|\mu+\nu|^2 = \left\{\begin{array}{ll}
\textrm{exp}(-2r), & \text{if } \phi_{\xi}=(2n+1)\pi\\
\frac{1}{2}\left[\textrm{exp}(2r)\left(1+\cos\phi_{\xi}\right)+\textrm{exp}(-2r)\left(1-\cos\phi_{\xi}\right)\right],  & \text{if } \phi_{\xi}\neq (2n+1)\pi,
\end{array}
\right.
\end{equation}
where $n=0, 1, 2, \ldots$. Now, by taking the first and the second case in the above equation as well as $r \rightarrow \infty$, we achieve the minimization and maximization, respectively. Both optimizations are shown in Figure \ref{fig1}. 

It should be noted that the displacements in the pointer [See Eqs. (\ref{position.shift}) and (\ref{momentum.shift})] reduce to
\begin{equation}
\delta x=gt\,\textrm{Re}(A_{w})\quad\text{and}\quad\delta p=\frac{2gt}{\hbar}\,\textrm{Var}(P)_{in}\,\textrm{Im}(A_{w})
\end{equation}
by means of the minimization. Furthermore, by choosing the optimization angle $\phi_{\xi}=2n\pi$ and $r\rightarrow\infty$, the pointer's shifts become
\begin{equation}
\delta x=gt\,\textrm{Re}(A_{c})\quad\text{and}\quad\delta p=0.
\end{equation}
Thus, the form of the Josza's theorem is recovered for the weak and strong regime for a suitable selection of squeezing parameters.

\section{\label{sec:formalism4}Trapped ion interacting with a bichromatic laser light}

\begin{figure}[h!]
	\centering
	\includegraphics[width=7.0cm, height=7.0cm]{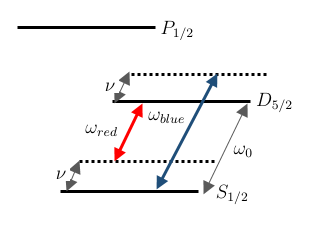}
	\caption{Configuration of a trapped ${}^{40}\text{Ca}^{+}$ ion with two internal states $S_{1/2}$ and $D_{5/2}$ whose transition frequency is $\omega_{0}$. The ion moves along the axial direction with a trapping frequency $\nu$ and interacts resonantly with a bichromatic laser of frequencies $\omega_{red}=\omega_{0}-\nu$ and $\omega_{blue}=\omega_{0}+\nu$. A third sublevel $P_{1/2}$ with lifetime $7.1\,\text{ns}$ is used to test the internal levels via resonance fluorescense.}
	\label{fig2}
\end{figure}

We will ilustrate the results obtained in the latter sections following the experimental setup shown in \cite{wu2019scheme}. In this work, they studied a single trapped ${}^{40}\text{Ca}^{+}$ ion inside the blade-shaped linear Paul trap, which allows it to oscillate along the axial direction with a frequency  $\nu=2\pi\times 1.41\,\text{MHz}$. The ion is considered as a two level system by taking into account the Zeeman sublevels $S_{1/2}\,(m_{J}=-1/2)$ and $D_{5/2}\,(m_{J}=1/2)$ that are identified as internal states $|\downarrow\rangle$ and $|\uparrow\rangle$, respectively. The transition between them is controlled by a narrow-linewidth laser at $729\,\text{nm}$. A bichromatic laser light interacting resonantly with the system causes the red and blue sidebands of the internal transition, which are driven by a acousto-optic modulator (See Figure \ref{fig2}). In the Lamb-Dicke regime \cite{javanainen1981laser}, this system is coupled to a pointer through the hamiltonian \cite{wallentowitz1995reconstruction, zheng1998preparation}
\begin{equation}
H=\eta\Omega\left(\sigma_{x}\sin\phi_{+}+\sigma_{y}\cos\phi_{+}\right)\otimes\left(X_{0}\sin\phi_{-} P-\frac{\hbar}{2X_{0}}\cos\phi_{-}X\right),
\end{equation}
where  $\Omega=2\pi\times 19\,\text{kHz}$ is the Rabi frequency, $\eta=0.08$ is the Lamb-Dicke parameter \cite{javanainen1981laser} and $\phi_{\pm}=\frac{1}{2}\left(\phi_{red}\pm\phi_{blue}\right)$ are phases related to the red sideband laser phase $\phi_{red}$ and the blue sideband laser phase $\phi_{blue}$ \cite{moya2012ion}. Here, $X=X_{0}\left(a+a^{\dagger}\right)$  and $P=\frac{\hbar}{2X_{0}} i \left(a^{\dagger}-a\right)$ are the position and momentum operator for the pointer in terms of the annihilation and creation operators. The motional state of the ion is characterized by a size $X_{0}=\sqrt{\hbar/(2m\nu)}=9.47\,\text{nm}$. In particular, by setting $\phi_{+}=\frac{\pi}{4}$ and $\phi_{-}=\frac{\pi}{2}$, the interaction hamiltonian takes the form
\begin{equation}
H=\gamma\,A\otimes P,
\end{equation}
where $\gamma=\eta\Omega X_{0}$, $A=\frac{1}{\sqrt{2}}\left(\sigma_{x}+\sigma_{y}\right)$ and the operator $P$ acts on a coherent squeezed state.

\begin{figure}[h!]
	\centering
	\includegraphics[width=8.6cm, height=11.5cm]{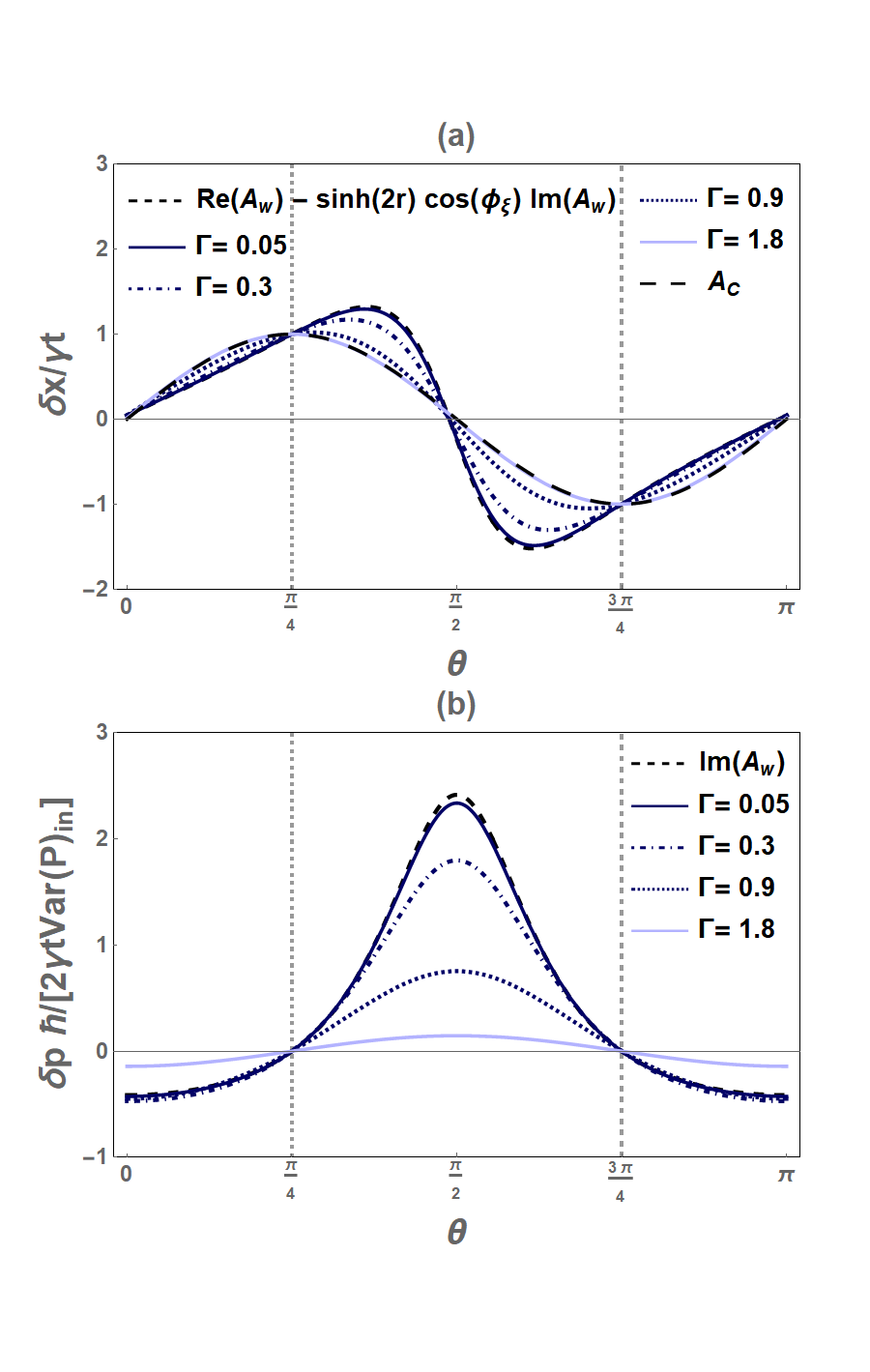}
	\caption{(Color online) The measurement transition of the pointer's shifts in the position (a) and the momentum (b) for a coherent squeezed initial pointer state after the post-selection of the system state by modifying the transition factor $\Gamma$. Here, we took $r=0.1$, $\phi_{\xi}=\frac{\pi}{6}$ and $\textrm{Im}\,\alpha=0.2$. The vertical dashed lines correspond to the eigenstate projections.}
	\label{fig3}
\end{figure}

\begin{figure}[h!]
	\centering
	\includegraphics[width=8.6cm, height=11.5cm]{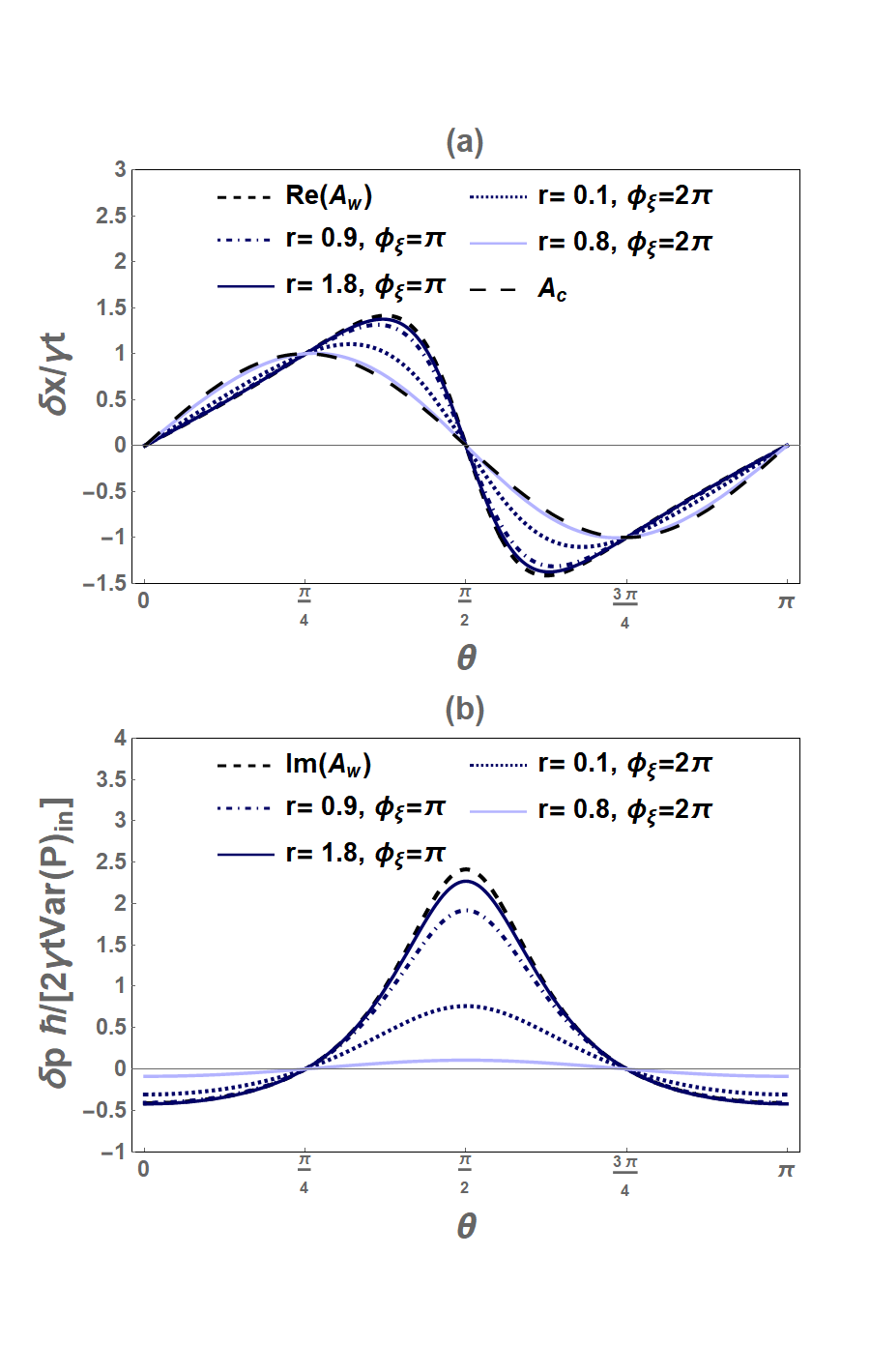}
	\caption{(Color online) Recovering the Josza's theorem by setting specific angles of squeezing $\phi_{\xi}$ for the pointer's shifts in the position (a) and the momentum (b) by varying the squeezing parameter $r$. Here, we chose $\textrm{Im}\,\alpha=0.01$ and $\Gamma=0.9$.}
	\label{fig4}
\end{figure}

A strategy to obtain the limits in the quantum measurement is only modulating the transition factor $\Gamma$ \cite{pan2020weak} for fixed squeezing parameters. In order to study the measurement transition, we consider the pre-selected and post-selected state as
\begin{equation}
|I\rangle=\frac{1}{\sqrt{2}}\left[|a_{+}\rangle+\textrm{exp}\left(i\frac{\pi}{2}\right)|a_{-}\rangle\right]
\end{equation} 
and
\begin{equation}
|F\rangle=\cos\theta\,|\uparrow\rangle+\textrm{exp}\left(i\frac{\pi}{4}\right)\sin\theta\,|\downarrow\rangle,
\end{equation}
where $|a_{\pm}\rangle=\frac{1}{\sqrt{2}}\left[|\uparrow\rangle\pm\textrm{exp}\left(i\frac{\pi}{4}\right)|\downarrow\rangle\right]$ are the eigenstates of $A$ whose eigenvalues are $\pm 1$, respectively. From Eqs. (\ref{position.shift}) and (\ref{momentum.shift}), it is possible to obtain the displacements of the pointer after the post-selection process by varying the transition factor $\Gamma$. For our system in study, this factor reduces to $\Gamma=\eta\Omega t$, which is proportional to the interaction time $t$. Hence, by changing this parameter we reach extreme regimes of the measurement. As shown in the Figure \ref{fig3}, there is a change in the shifts of the pointer from the weak regime $(\Gamma = 0.05)$ to the strong regime $(\Gamma=1.8)$, where we used parameters compatible with the experimental work in Ref. \cite{pan2020weak}. It should be emphasized that the form of the Josza's theorem \cite{jozsa2007complex} is not regained in the spatial displacement for a coherent squeezed pointer state. However, it can be recovered by taking $r=0$ (coherent pointer state) or $\phi_{\xi}=n\pi\,(n=0,1,2,\ldots)$. As we have seen in Section \ref{sec:formalism3}, by choosing $r \rightarrow \infty$ and $\phi_{\xi}=(2n+1)\pi$, the displacements in the pointer depend on the real and imaginary part of the weak value. Similarly, if we take $\phi_{\xi}=2n\pi$, the conditional expectation value is related to the pointer's shifts (See Figure \ref{fig4}).   

\begin{figure}[h!]
	\centering
	\includegraphics[width=8.6cm, height=11.5cm]{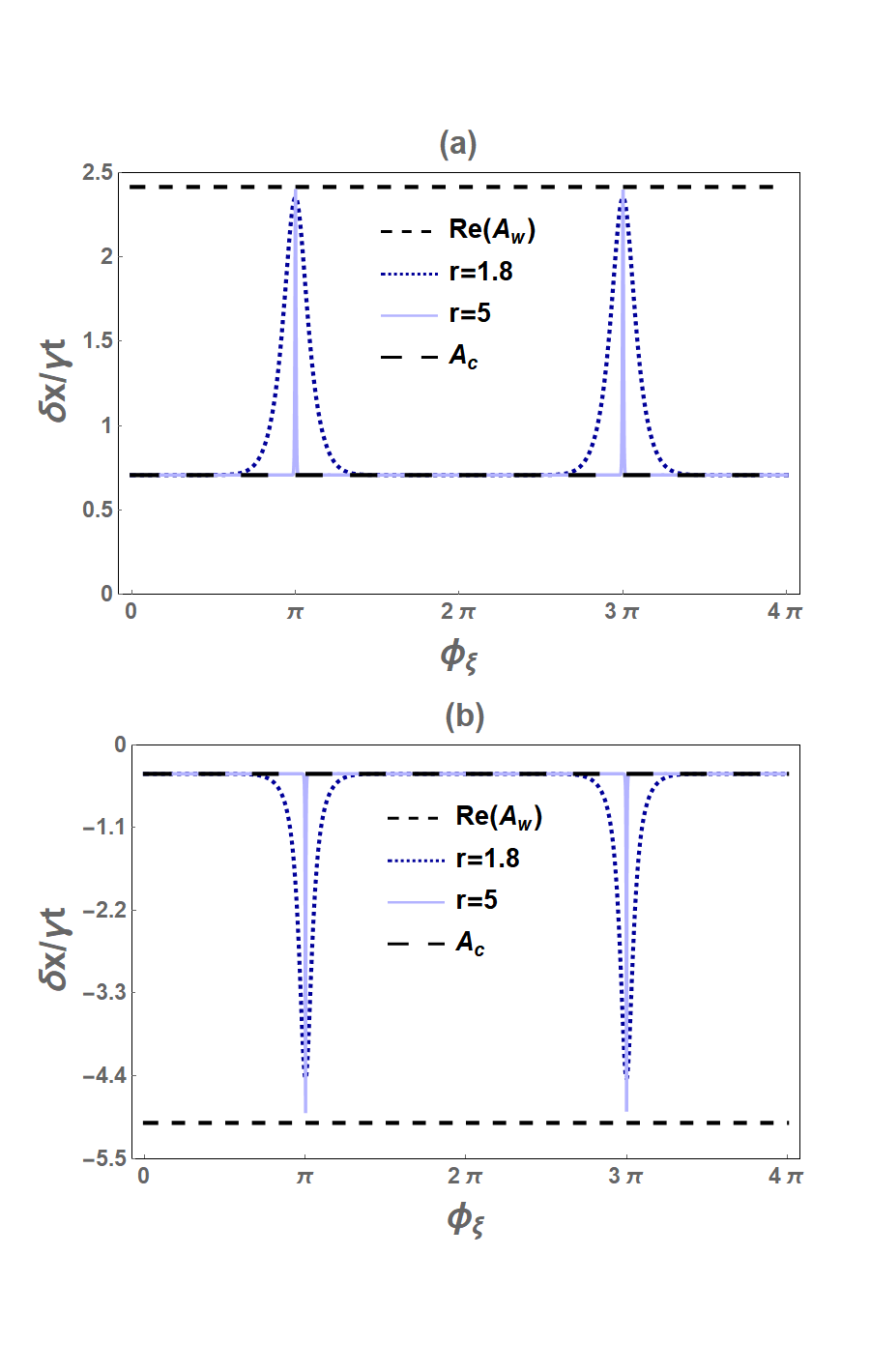}
	\caption{(Color online) A useful strategy to obtain the measurement transition from the weak to the strong regime by setting the following post-selection angles: (a) $\vartheta=\frac{\pi}{8}$ and (b)  $\vartheta=\frac{5\pi}{16}$. Here, we took $\textrm{Im}\,\alpha=0$ and $\Gamma=0.9$.}
	\label{fig5}
\end{figure}

 Clearly, the measurement transition can be achieved by only tuning the squeezing parameters $r$ and $\phi_{\xi}$. Now, we show how these parameters influence in a dramatic way the spatial displacement by choosing the following states
\begin{equation}
|I\rangle=\frac{1}{\sqrt{2}}\left(|a_{+}\rangle-|a_{-}\rangle\right) \quad \text{and} \quad |F\rangle=\cos\vartheta\,|a_{+}\rangle+\sin\vartheta\,|a_{-}\rangle.
\end{equation}
By regarding these states and $\textrm{Im}\,\alpha=0$, the transition value is real implying a change only in the pointer's position [See. Eqs (\ref{position.shift}) and (\ref{momentum.shift})]. The Figure \ref{fig5} shows our strategy to cause the measurement transition by maintaining the global transition factor and adjusting the squeezing parameters. For the system in study, we took a fixed global transition factor $\Gamma=0.9$ considering a specific interaction time, which is consistent with the experimental scheme shown previously. Notice that for a large squeezing parameter $r$, the figure shows that the pointer's displacement goes from the conditional expectaction value to the weak value, in sharp peaks, for $\phi_{\xi}=(2n+1)\pi$. 

The trapped ion system and generation of squeezed phonons was chosen motivated by the experiment reported in Ref. \cite{pan2020weak,kienzler2015quantum, drechsler2020state}. although our calculations are applicable to other systems as well \cite{meekhof1996generation, burd2019quantum}.



\begin{figure}[h!]
	\centering
	\includegraphics[width=8.6cm, height=6.5cm]{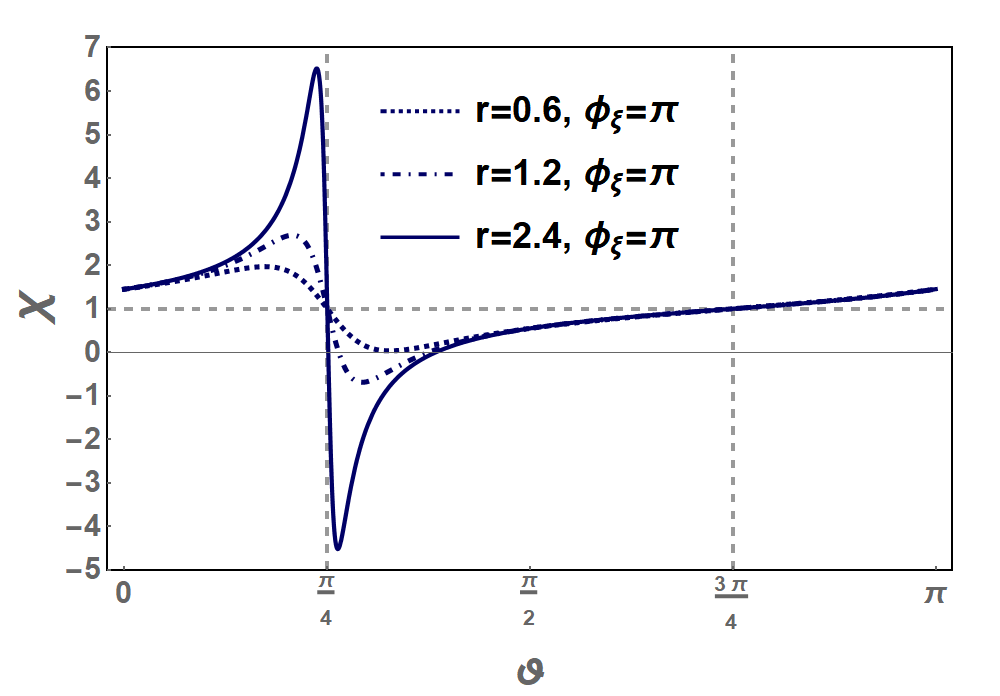}
	\caption{(Color online) A method to control amplification by leading the quantum measurement to the weak regime $[r \rightarrow \infty\;\text{and}\; \phi_{\xi}=(2n+1)\pi]$ and modifying the post-selection angle $\vartheta$ from $\vartheta=3\pi/4$ where does not produce amplification to reaching a maximum at $\vartheta\approx \pi/4$ (WVA). The dimensionless gain factor $\chi=X_{fin}/X_{in}$ is the ratio between the pointer's position after the post-selection process $X_{f}=\langle \phi_{fin}|X|\phi_{fin}\rangle/\langle \phi_{fin}|\phi_{fin} \rangle$ and the initial position of the pointer $X_{in}=\langle \phi_{in}|X|\phi_{in} \rangle$. This factor describes quantitatively  the amplification of the position in the pointer caused by the measurement. Here, we selected $\Gamma=0.9$ and $\alpha=1$.}
	\label{fig6}
\end{figure}

\section{Conclusions}

In summary, we find the pointer's shifts in the position and momentum for a coherent squeezed pointer state by using an arbitrary measured observable $A$. These expressions are linked to the real and imaginary part of the transition value which is defined in this work. By modulating the transition factor $\Gamma$, expressions for different measurement regimes are obtained that generalize the coherent and ground pointer state \cite{turek2015advantages,zhu2011quantum}. Besides, by choosing certain squeezing angles $\phi_{\xi}$ and taking the parameter $r\rightarrow \infty$, Josza's results are recovered \cite{jozsa2007complex}. We also present a new strategy to reach the weak and strong regime by only modifying the squeezing parameters $r$ and $\phi_{\xi}$. All these ideas have been inspired by the experiments on Calcium ion \cite{pan2020weak,kienzler2015quantum, drechsler2020state}. The results in Figure \ref{fig5} show that by varying the squeezing phase, one can achieve a fast transition from the strong to weak regime. In case we set the paramenters as to have weak value amplification, we are seeing a fast transition from the no amplification $(\chi=1)$ to a possible maximal amplification $(\chi \rightarrow \pm \infty)$ under special conditions of the post-selection angle as ilustrated in Figure \ref{fig6}, which may lead to some interesting physical applications such as an amplification regulator of signals where small signals, equivalent to small displacements of the pointer, are detected by a measuring device and they can be amplified by choosing $r \rightarrow \infty\;\text{and}\; \phi_{\xi}=(2n+1)\pi$. Then by tuning the post-selection angle the maximal amplification (WVA) is achieved when $\langle  F|I \rangle \rightarrow 0$. This method allows the amplification of signals by only using squeezing parameters of the motion state as well as a suitable post-selection process.


\begin{acknowledgments}
	We thank to the FONDECYT project \#1180175 and Beca Doctorado Nacional ANID \#21181111, for financial support.
\end{acknowledgments}

\appendix*
\section{Proof of some useful expressions}
\begin{enumerate}
	\item The inner product between coherent squeezed states is
	\begin{equation}\label{inner.product.squeezed.state}
	\langle y, \xi|z, \xi \rangle=\textrm{exp}\left[i\,\textrm{Im}(y^{*}z)\right]\,\textrm{exp}\left(-\frac{1}{2}\,|\delta|^2\right).
	\end{equation} 
	Proof.
	
	First, by using the definition of the coherent squeezed state [See. Eq. (\ref{squezeed.state})] and the property in Eq. (\ref{coherent.property}) results
	\begin{eqnarray}\label{eq1.app}
	\langle y, \xi|z, \xi \rangle &=& \langle 0|S^{\dagger}(\xi)D^{\dagger}(y)D(z)S(\xi)|0\rangle\nonumber\\
	&=&	\langle 0|S^{\dagger}(\xi)D(-y)D(z)S(\xi)|0\rangle\nonumber\\
	&=&\textrm{exp}\left[i\,\textrm{Im}\left(y^{*}z\right)\right]\langle 0|S^{\dagger}(\xi)D(z-y)S(\xi)|0\rangle
	\end{eqnarray}
	Then by using the properties \cite{orszag2016quantum}
	\begin{equation}
	D(b)S(\xi)=S(\xi)D(c),
	\end{equation}
	where $c=\mu b+\nu b^{*}$ and $\langle b|c \rangle=\textrm{exp}\left[-\frac{1}{2}\left(|b|^2+|c|^2-2b^{*}c\right)\right]$, Eq. (\ref{eq1.app}) reduces to
	\begin{eqnarray}
	\langle y, \xi|z, \xi \rangle &=&\textrm{exp}\left[i\,\textrm{Im}\left(y^{*}z\right)\right]\langle 0|D(\delta)|0\rangle\nonumber\\
	&=&\textrm{exp}\left[i\,\textrm{Im}\left(y^{*}z\right)\right]\langle 0|\delta\rangle\nonumber\\
	&=&\textrm{exp}\left[i\,\textrm{Im}\left(y^{*}z\right)\right]\,\textrm{exp}\left(-\frac{1}{2}\,|\delta|^2\right)
	\end{eqnarray}
	with 
	$\delta=\mu(z-y)+\nu(z^{*}-y^{*})$. Here $\mu=\cosh{r}$ and $\nu=\textrm{exp}\left(i\phi_{\xi}\right)\,\sinh{r}$.
	
	\item An application of the inner product between coherent squeezed states is
	\begin{equation}\label{special.inner.product}
	\langle y, \xi|a|z, \xi\rangle=\left[\mu^{2} z-|\nu|^2 y+\mu\nu(z^{*}-y^{*})\right]\langle y, \xi|z, \xi \rangle.
	\end{equation}
	Here, $a$ is the annihilation operator.
	
	Proof.
	
	Firstly, we write the annihilation operator in terms of a generalized annihilation operator $A$ \cite{orszag2016quantum} as
	\begin{equation}
	a=\mu A-\nu A^{\dagger}.
	\end{equation}
	Then 
	\begin{equation}\label{eq2.app}
	\langle y, \xi|a|z, \xi\rangle=\mu \langle y, \xi|A|z, \xi\rangle-\nu \langle y, \xi|A^{\dagger}|z, \xi\rangle.
	\end{equation}
	Finally, with the help of the following property \cite{orszag2016quantum}
	\begin{equation}
	A|b,\xi\rangle=c\,|b,\xi\rangle,
	\end{equation}
	where $c=\mu b+\nu b^{*}$, Eq. (\ref{eq2.app}) becomes
	\begin{equation}
	\langle y, \xi|a|z, \xi\rangle=\left[\mu^{2} z-|\nu|^2 y+\mu\nu(z^{*}-y^{*})\right]\langle y, \xi|z, \xi \rangle.
	\end{equation}
\end{enumerate}

\bibliographystyle{apsrev4-1}
\bibliography{biblio}

\end{document}